\documentclass[twocolumn]{svjour3}          
\smartqed  
\usepackage{graphicx}
\usepackage[utf8]{inputenc}
\usepackage{amsmath,amssymb,latexsym}
\usepackage{cite}

\newcommand{\htwo}{\ensuremath{\text{H}_2}}
\newcommand{\heh}{\ensuremath{\text{HeH}^+}}
\newcommand{\lih}{\ensuremath{\text{LiH}}}

\DeclareMathOperator*{\argmin}{arg\,min}

\def\makeheadbox{{%
\hbox to0pt{\vbox{\baselineskip=10dd\hrule\hbox
to\hsize{\vrule\kern3pt\vbox{\kern3pt
\hbox{\bfseries [International Journal of Dynamics and Control]}
\hbox{This is a pre-acceptance version of an article currently in the peer-review process.}
\kern3pt}\hfil\kern3pt\vrule}\hrule}%
\hss}}}

\begin{document}

\title{Machine Learning a Molecular Hamiltonian for Predicting Electron Dynamics}



\author{Harish S. Bhat         \and
        Karnamohit Ranka \and 
        Christine M. Isborn 
}


\institute{Harish S. Bhat \at
              Applied Mathematics Department\\
              University of California, Merced\\
              5200 N. Lake Rd., Merced, CA 95343\\
              \email{hbhat@ucmerced.edu}           
           \and
           Karnamohit Ranka \at
              Chemistry Department\\
              University of California, Merced\\
              5200 N. Lake Rd., Merced, CA 95343\\
              \email{kranka@ucmerced.edu}  
           \and
           Christine M. Isborn \at
              Chemistry Department\\
              University of California, Merced\\
              5200 N. Lake Rd., Merced, CA 95343\\
              \email{cisborn@ucmerced.edu}           
}

\date{Received: date / Accepted: date}

\maketitle

\begin{abstract}
We develop a computational method to learn a molecular Hamiltonian matrix from matrix-valued time series of the electron density.  As we demonstrate for three small molecules, the resulting Hamiltonians can be used for electron density evolution, producing highly accurate results even when propagating $1000$ time steps beyond the training data.  As a more rigorous test, we use the learned Hamiltonians to simulate electron dynamics in the presence of an applied electric field, extrapolating to a problem that is beyond the field-free training data.  We find that the resulting electron dynamics predicted by our learned Hamiltonian are in close quantitative agreement with the ground truth.  Our method relies on combining a reduced-dimensional, linear statistical model of the Hamiltonian with a time-discretization of the quantum Liouville equation within time-dependent Hartree Fock theory.  We train the model using a least-squares solver, avoiding numerous, CPU-intensive optimization steps.  For both field-free and field-on problems, we quantify training and propagation errors, highlighting areas for future development.

\end{abstract}

\section{Introduction}
\label{intro}
An intriguing new application of machine learning is to predict the dynamical electronic properties of a molecular system \cite{Hase2016, Gastegger2017, Chen2018}, which is essential to understanding phenomena such as charge transfer and response to an applied laser field.  When discussing such electron dynamics predictions, we must start with the electronic time-dependent Schr\"odinger equation (TDSE):
\begin{equation}
i\frac{d\Psi(\mathbf{r},t)}{dt} = \hat H (\mathbf{r},t) \Psi(\mathbf{r},t). \label{eq:TDSE}
\end{equation}
Here $\hat H(\mathbf{r},t)$ is the electronic Hamiltonian operator that operates on the time-dependent many-body electronic wave function $\Psi(\mathbf{r},t)$, where $\mathbf{r}$ represents the spatial and spin coordinates of all electrons.   One can derive from (\ref{eq:TDSE}) an evolution equation for the time-dependent density operator.  This operator equation can be represented in a finite-dimensional basis, yielding a matrix system of ordinary differential equations:
\begin{equation}
\label{eq:TDHF_density}
i\frac{d \mathbf{P'}(t)}{d t} =  \bigg [   \mathbf{H'}(t), \mathbf{P'}(t) \bigg].
\end{equation}
We call this the quantum Liouville-von Neumann equation.  Boldface capital letters denote matrices, representations of operators in particular bases. Primes denote representations of operators in an orthonormal basis.  Here $\mathbf{P'}(t)$ and $\mathbf{H'}(t)$ are time-dependent density and Hamiltonian matrices, respectively. The square brackets on the right-hand side denote a commutator; for matrices $\mathbf{A}$ and $\mathbf{B}$, the commutator is $[\mathbf{A}, \mathbf{B}] = \mathbf{A} \mathbf{B} - \mathbf{B} \mathbf{A}$.

Inspecting a particular molecular system, one determines and writes the system's Hamiltonian, a sum of kinetic and potential energy operators.  As the Hamiltonian includes spatial derivatives within the kinetic energy operator, (\ref{eq:TDSE}) will be a partial differential equation (PDE).  For an $N$-electron system, ignoring spin, the PDE (\ref{eq:TDSE}) will feature $3N$ spatial degrees of freedom.  As $N$ increases beyond $N = 1$, it becomes intractable to solve (\ref{eq:TDSE}) directly for the time-dependent many-body wave function $\Psi(\mathbf{r},t)$, even with modern numerical analysis and high-performance computing. For this reason, molecular electronic structure and dynamics calculations typically use simplified, mean-field approaches.  One such approach is time-dependent Hartree-Fock (TDHF)\cite{Runge94_322,HeadGordon05_4009} theory, which solves (\ref{eq:TDHF_density}) based on a simplified form of the wave function.  In HF theory, we approximate the many-body wave function using a Slater determinant, an anti-symmetrized product of single-particle orbitals $\phi_i (\mathbf{r}, t)$, where $\mathbf{r}$ now represents the spatial and spin coordinates of one electron.  This approximation leads to a modified form of the Hamiltonian $\hat H$ that appears in (\ref{eq:TDSE})---for details, see (\ref{eq:nonrelativisticH}) below. Within HF theory, we then call (\ref{eq:TDHF_density}) the TDHF equation. 

Equation (\ref{eq:TDHF_density}), used within TDHF theory or an alternative, similar approach called time-dependent density functional theory, is used in atomic, molecular, and materials calculations to simulate the dynamic electronic response to a perturbation, including predicting charge transfer and spectroscopic properties \cite{Li2005,Li07_134307,Manby08_114113,Govind11_1344,Isborn16_739,Herbert18_044117,DePrince16_5834}. 
In these physical science settings, one starts with a system of interest, e.g., a molecule in an applied electric field.  The system's atomic configuration completely determines the Hamiltonian $\mathbf{H'}$ and therefore the right-hand side of (\ref{eq:TDHF_density}).  Starting from an initial condition, the typical workflow is then to numerically solve (\ref{eq:TDHF_density}) forward in time to generate simulations of interest, i.e., to generate $\mathbf{P'(t)}$ for $t > 0$ for a given perturbation.

Note that we write $\mathbf{H}'(t)$ to encapsulate two types of dependence on time $t$.  First, $\mathbf{H}'$ can depend explicitly on time---we will see this below when we consider $\mathbf{V}_{\mathrm{ext}}$, an external, time-dependent potential.  Second, within HF theory, even if $\mathbf{H}'$ does not depend explicitly on time, it \emph{is} in general a function of the density $\mathbf{P}'(t)$.  In summary, $\mathbf{H}'(t)$ is shorthand for $\mathbf{H}'(t, \mathbf{P}'(t))$.  This implies that (\ref{eq:TDHF_density}) is in fact a \emph{nonlinear} system.

\paragraph{Main Contribution.} In this paper, we address a \emph{system identification} problem for (\ref{eq:TDHF_density}).  Our main contribution is a computational method to estimate the molecular field-free matrix Hamiltonian $\mathbf{H'}(t)$ from time series observations of density matrices $\mathbf{P'}(t)$.  By building a data-driven model of $\mathbf{H}'$, we identify the right-hand side of (\ref{eq:TDHF_density}).   We use a linear model for $\mathbf{H}'$,  formulate a quadratic loss function that stems from discretizing (\ref{eq:TDHF_density}) in time, and  eliminate unnecessary degrees of freedom.  Thus we reduce model training to a least-squares problem.  We demonstrate this method using training data consisting of density matrices $\mathbf{P}(t)$ for three small molecules.

Among other tests, we use the machine-learned (ML) Hamiltonian to propagate, i.e., to solve (\ref{eq:TDHF_density}) forward in time.  We find that using the ML Hamiltonian instead of the exact Hamiltonian results in a small, acceptable level of propagation error, even on a time interval that is twice the length of the time interval used for training.  We then add a time-dependent external potential to our machine-learned, field-free Hamiltonian; we propagate forward in time using this augmented Hamiltonian.  For each of the three molecules we consider, the resulting solutions are in close quantitative agreement with simulations that use the exact Hamiltonian.  In short, our machine-learned Hamiltonian extrapolates well to a dynamical setting that differs from that of the training data.

To our knowledge, despite the surge of interest in applying machine learning to molecular simulation \cite{Snyder2012,Montavon2013,Ramakrishnan2015,Bartok2017,Grisafi2018,Nebgen2018,Paruzzo2018,Pronobis2018,Sifain2018,Rodriguez2019,Christensen2019,Ghosh2019,Wilkins2019,Ye2019,Chandrasekaran2019,Schleder2019,Jorgensen2019,Smith2019,Ceriotti2019,Lu2020}, there are no other procedures in the literature to estimate molecular Hamiltonians from density matrix time series.  Our work shares goals with other efforts to learn Hamiltonians, or energy functions and functionals that are ingredients in Hamiltonians.  In this space, we primarily see efforts to learn classical Hamiltonians from time series \cite{Bertalan2019, Bhat2019, Rezende2019, Chen2020, Jin2020, Miller2020, Protopapas2020, Toth2019, Zhong2020} as well as efforts to learn quantum Hamiltonians or potentials for time-independent problems \cite{Behler2007, Behler2016, LiCollins2018, Fujita2018, Innocenti2020}.  Recently, a neural network method to learn the exchange-correlation functional in time-dependent density functional theory has been developed \cite{Suzuki2020}; solutions of the corresponding TDSE are used to train the networks.

 In this paper, we consider small molecular systems modeled with a small basis set in order to focus on methodological development and careful analysis of errors.  The present work forms a foundation on which we can build towards studying systems and theories (such as time-dependent density functional theory) in which the underlying potentials and functionals have yet to be completely determined.  This is the overarching motivation for pursuing the present work.

\paragraph{Extrapolation.} To clarify what we mean by extrapolation, let us consider a classical mass-spring system; one end of the spring is fixed and the other is connected to the mass.  Let $F(t)$ be an external force applied to the mass, and let $x(t)$ denote the displacement from equilibrium of the mass at time $t$.  We start with time series of $x(t)$ for a system where $x(0) \neq 0$ but $F(t) \equiv 0$, i.e., a field-free system.  With this training data, suppose we seek a machine learning model that can predict the mass' position accurately, even when we switch on an applied forcing, e.g., $F(t) = A \cos(\omega t)$.

We view the task of (i) training with field-off data and (ii) predicting for field-on systems as an extrapolation task.  For this classical example, one approach is to learn a model of $V(x)$, the spring's potential.  Equipped with a sufficiently accurate model of $V$, the field-free dynamics can be predicted from Newton's second law: $m \ddot{x} = -V'(x)$.  Again assuming that $V$ is sufficiently accurate, we can predict the behavior of the mass-spring system accurately when the field is on, by solving $\ddot{x} = -V'(x) + F(t)$.  Suitably expanded in size and complexity to the TDHF equation (\ref{eq:TDHF_density}), this is essentially what we do in the present work. 




\section{Physical Considerations}
\label{sect:physics}
\subsection{Time-Dependent Hartree-Fock}
In Section \ref{intro}, we provided a highly summarized conceptual overview of deriving the TDHF equation from the TDSE.  Here we expand on this overview and give more mathematical details.  
All equations use atomic units, with $e^2 = \hbar = m_e = 1$. An external perturbation, such as an applied electric field, within the Hamiltonian will give rise to the time-evolution of the wave function that dictates all properties of a quantum electronic system. 

The molecules studied are closed-shell systems---all electrons in the system are spin-paired. Within HF theory, each pair of electrons with opposite spins can be described by the same spatial function $\phi_i$, which is referred to as \textit{restricted} HF theory. For closed-shell systems within such a formalism, the need to solve for $N$ spatial orbitals occupied by $N$ electrons reduces to solving for $(N/2)$ spatial orbitals, each of these doubly occupied to give a total of $N$ electrons \cite{SzaboOstlund1996}.

Given this choice, our Hamiltonian is then
\begin{multline}
\hat H(\mathbf{r},t) =\sum_{i}^{N/2}\left[ -\frac{\nabla^2}{2} - \sum_A \frac{Z_A}{| \mathbf{r} - \mathbf{R}_A|} +  \hat{V}_{\mathrm{ext}}(\mathbf{r},t) \right] \\
\qquad + \sum_{i}^{N/2} \bigg [ \sum_j^{N/2} \ 2 \times \int d\mathbf{r}' \frac{\phi_j^*(\mathbf{r}', t)\phi_j(\mathbf{r}', t)}{| \mathbf{r} - \mathbf{r}'| }    \\
- \int d\mathbf{r}' \frac{\phi_j^*(\mathbf{r}', t) \mathfrak{P} (\mathbf{r},\mathbf{r}') \phi_j(\mathbf{r}', t)}{| \mathbf{r} - \mathbf{r}'| } \bigg ],
\label{eq:nonrelativisticH}
\end{multline}
where the first group includes one-electron terms summed over half the number of electrons (the total number of \textit{spatially unique} electrons): the electron kinetic energy, the electron-nuclear attraction to all nuclei $A$ with nuclear charge $Z_A$ at fixed position $\mathbf{R}_A$, and the external potential $\hat{V}_{\mathrm{ext}}$. In this work, the external perturbation is an electric field treated classically within the dipole approximation $\hat{V}_{\mathrm{ext}}(\mathbf{r},t) = \mathbf{E} (t)\cdot \hat{\mu} (\mathbf{r})$. The first term in the second group is the electron-electron Coulomb repulsion. The operator $\mathfrak{P}(\mathbf{r}$,$\mathbf{r}')$ used in the last term denotes the operation of permutation between electrons represented by coordinate-variables $\mathbf{r}$ and $\mathbf{r}'$. This term is known as the exchange operator and arises from the antisymmetry of the electronic wave function. The second group collectively represents the electron-electron interaction operator. This operator depends on the instantaneous charge distribution of all other electrons, resulting in an implicit time-dependence of the electronic Hamiltonian (in addition to the explicit time-dependence due to $\hat{V}_{\mathrm{ext}}$).

We next define a (reduced one-body) density operator, $\hat{\rho}$, that allows us to represent the total density of electrons\cite{Dirac1930}:
\begin{equation}
    \hat{\rho} (\mathbf{r}, t) = \sum_p f_p \phi_p (\mathbf{r}, t) \phi_p^* (\mathbf{r}, t) 
    = \sum_p f_p | \phi_p \rangle \langle \phi_p |,
\end{equation}
where $f_p$ is the occupation of orbital $\phi_p$: in a restricted, closed-shell system, $f_p$ = 2 (if $\phi_p$ is occupied) or 0 (if $\phi_p$ is unoccupied). The corresponding density matrix ($ \mathbf{P} $) is represented in the basis of $ \left\{ \phi_i \right\} $ as:
\begin{equation}
    P_{ij} (t) = \! \int \! d\mathbf{r} \phi_i^* (\mathbf{r}, t) \hat{\rho} (\mathbf{r}, t) \phi_j (\mathbf{r}, t) = \langle \phi_i | \hat{\rho} | \phi_j \rangle.
\end{equation}
We can now write down the Liouville-von Neumann equation in operator form: 
\begin{equation}
i\frac{d\hat{\rho}(\mathbf{r}, t)}{dt} = [ \hat{H}(\mathbf{r}, t), \hat{\rho}(\mathbf{r}, t) ]. 
\end{equation}
This is an operator equation for the evolution of $\hat{\rho}$.  The time-dependent molecular orbitals $\phi_i$ are often created from a linear combination of basis functions $\{\chi_\mu\}$, as $\phi_i = \sum_\mu c_{\mu, i}(t)\chi_\mu $, where $c_{\mu, i}(t)$ are the time-dependent coefficients. The elements of the density matrix $\mathbf{P}$ are given in this basis by 
\begin{equation} \label{eq:dens_AO}
P_{\mu \nu}(t) = \sum_p f_p c_{\mu, p}(t) c^*_{\nu, p}(t).  
\end{equation}
We transform \textbf{P} to an orthonormal basis, yielding  $\mathbf{P'}$ (see Appendix).  We then use the Liouville equation for the density operator to write the TDHF equation in matrix form
\begin{equation}
\label{eq:TDHF_density2}
i\frac{d \mathbf{P'}(t)}{d t} =  \bigg [   \mathbf{H'}(t), \mathbf{P'}(t) \bigg],
\end{equation}
where $\mathbf{H'}(t)$ is the Hamiltonian (or Fock) matrix that results from integrating (\ref{eq:nonrelativisticH}) over $\mathbf{r}$ in the orthonormal basis. In this work, primed notations (\emph{e.g.}, $\mathbf{H}',\mathbf{P}'$) are used for matrices in the orthonormal basis and unprimed notations for matrices (\emph{e.g.}, $\mathbf{H},\mathbf{P}$) in the atomic orbital (AO) basis.

Although it is straightforward to write down the molecular Hamiltonian if the atomic positions are known, integration of the Hamiltonian within a given basis is more challenging and encodes ground and excited state information about the molecule within that basis. Learning the integrated form of the molecular matrix Hamiltonian is thus key to determining the electron dynamics. 

\subsection{Molecules and Exact Hamiltonian}
Here we study three diatomic molecules: \htwo, \heh, and \lih. The atoms in each of these diatomic systems are placed along the $z$-axis, equidistant from the origin. The interatomic separations for \htwo, \heh\ and \lih\ are 0.74 \r{A}, 0.772 \r{A}, and 1.53 \r{A}, respectively. These simple molecular systems increase in complexity, going from a symmetric two-electron homonuclear diatomic, to a two-electron heteronuclear diatomic, to a four-electron heteronuclear diatomic.  The basis set used for these calculations is STO-3G, a minimal basis set made of $s$ and $p$ atomic orbitals. For \htwo\, and \heh, this results in two basis functions (a $2 \times 2$ matrix for $\mathbf{P}$ and $\mathbf{H}$), and for LiH this results in six basis functions (a $6\times6$ matrix for $\mathbf{P}$ and $\mathbf{H}$, although some elements of the matrices are zero due to the linear symmetry of the molecule, as discussed later). 

For each molecule, the electronic structure code provides the integrals that are the components of the exact Hamiltonian matrix $\mathbf{H}$, expressed in the same AO basis set as the density matrices. Specifically, we obtain real, symmetric, constant-in-time matrices for the kinetic energy and electron-nuclear potential energy.  We also obtain a $4$-index tensor of evaluated integrals, which we use together with the time-dependent density matrices $\mathbf{P}(t)$ to compute the electron-electron potential energy term.  These ingredients allow us to compute, for each molecule, the exact Hamiltonian. Electron density propagation with this exact Hamiltonian, both within the electronic structure code and within our propagation code, is compared to that from our ML model Hamiltonian. 

\section{Electron Density Matrix Data}
\label{sec:1}
There are two steps involved in generating the training and test sets of the time series of density matrix data:
\begin{enumerate}
    \item Generating an initial condition (the initial density matrix).
    \item Generating a trajectory using the initial condition and the differential equation \eqref{eq:TDHF_density} for propagation.
\end{enumerate}
For the first step, the HF stationary state solution is determined self-consistently within the electronic structure code. The density matrix corresponding to the alpha-spin part of the solution, represented in the AO basis, is used as the initial condition. The second step involves propagating the initial density matrix using the TDHF equation.

We performed each of these steps with the Gaussian electronic structure program\cite{GaussianDV}, using a locally modified development version.

\subsection{Initial Conditions}
\label{sect:initcond}
We have calculated initial density matrices for field-free and static field conditions. 
For the field-free calculations, we set the $\mathbf{V}_{\mathrm{ext}}$ term to 0. For the static field, $E_z$ = 0.05 a.u. (atomic units). Applying a static field creates an initial electron density that is not a stationary state of the field-free Hamiltonian and is often referred to as a delta-kick perturbation. 

\subsection{Trajectory Data}
\label{sect:trajdata}
The density matrix from the initial condition calculation is used as the starting point for generating the real-time TDHF electron dynamics trajectory, i.e. $\mathbf{P}(t)$.

For the field-free trajectories, we set
$\mathbf{V}_{\mathrm{ext}}$ to zero during propagation; we use the density matrix with the delta kick perturbation as the initial condition. These trajectories serve as the training data for the ML Hamiltonian.  There is a particular rationale behind choosing the delta kick perturbation.  First, consider that a perturbation that is localized at one point in time is, by Fourier duality, maximally spread out in frequency space.  Hence such a perturbation necessarily excites all modes of the system.  Second, consider that if we generate a trajectory that does not excite a mode of a system, we cannot expect that a machine learning technique will be able to learn that that mode exists, let alone how to represent the unobserved mode accurately in either the potential or the full Hamiltonian.  By choosing a delta kick trajectory that excites all modes, we ensure that it is at least possible in principle to learn the full potential/Hamiltonian.

For the field-on trajectories, the field-free initial density matrix is used and $\mathbf{V}_{\mathrm{ext}}$ takes the following form during  propagation:
\begin{equation}
\label{eqn:Efield}
    \mathbf{V}_{\mathrm{ext}}(t) = \! \! \! \!  \! \! \sum\limits_{i \in \{ x,y,z \}}  \! \! \!  \! \! \! E_i \sin{(\omega  t)}  \boldsymbol \mu_i = 0.05 \sin{(0.0428  t)} \boldsymbol \mu_z,
\end{equation}
where the time $t$, the field-intensity $E_i$ along axis $i$, and the field-frequency $\omega$ are expressed in a.u. Here the field is applied only along the z-direction and $\boldsymbol {\mu_z}$ is the z-component of the dipole moment matrix in the AO basis. The sinusoidal field is switched on for one full cycle (around 3.55 fs) starting at $t=0$. These field-on trajectories test the ML Hamiltonian in a regime quite outside the field-free training regime.  

Using a propagation step-size of 0.002 fs, the total length of each trajectory is 20000 time-steps (thus, each trajectory is 40 fs long).  The real-time TDHF implementation in Gaussian uses as its propagation scheme the modified midpoint unitary transformation (MMUT) algorithm\cite{Li2005}.

\section{Learning the Molecular Hamiltonian}
\label{sec:learnham}

For a particular molecule, suppose we are given time series $\{ \mathbf{P'}(t_j) \}_{j=0}^N$ sampled on an equispaced temporal grid $t_j = j \Delta t$.  We assume that $\mathbf{P'}(t)$, the continuous-time trajectory  corresponding to our time series, satisfies (\ref{eq:TDHF_density}). Our goal is to learn the Hamiltonian $\mathbf{H}'$.  Assume that the Hamiltonian contains no explicit time-dependence---this can be ensured by generating training data with no external potentials (e.g., no external forcing).  Then $\mathbf{H}'$ is a Hermitian matrix of functions of $\mathbf{P}'$, the density matrix.  Our strategy therefore consists of three steps: (i) develop a model of $\mathbf{H}'$ with a finite-dimensional set of parameters $\beta$, (ii) derive from (\ref{eq:TDHF_density}) a statistical model, and (iii) use the model with available data to estimate $\beta$.

Note that in order to obtain $\mathbf{P}', \mathbf{H}'$ from $\mathbf{P}, \mathbf{H}$, we transform from the AO basis to its canonical orthogonalization \cite{SzaboOstlund1996}. We do this because the TDHF equation (\ref{eq:TDHF_density}) holds in an orthonormal basis; the AO basis by itself is not orthonormal.  We leave the details of this transformation to the Appendix.

Let us split $\mathbf{H}'$ into its real and imaginary parts: $\mathbf{H}' = \mathbf{H}'_R + i \mathbf{H}'_I$.  By Hermitian symmetry, $\mathbf{H}'$ is determined completely by the upper-triangular component of $\mathbf{H}'_R$ (including the diagonal) and by the upper-triangular component of $\mathbf{H}'_I$ (not including the diagonal).  If $\mathbf{H}'$ has size $M \times M$, there are $M(M+1)/2$ elements of $\mathbf{H}'_R$ and $M(M-1)/2$ elements of $\mathbf{H}'_I$ that we must model.  Hence there are a total of $M^2$ real degrees of freedom, which we can represent as an $M^2 \times 1$ vector $\mathbf{h}'$.  Note that we can apply this same real and imaginary splitting to $\mathbf{P}'$; since it is also Hermitian, it can also be determined completely by a real vector $\mathbf{p}'$ of dimension $M^2 \times 1$.  Then we formulate the following linear model for $\mathbf{h}'(\mathbf{p}')$---in what follows, we use $\widetilde{\ }$ to denote either statistical models or their parameters:
\begin{equation}
    \label{eqn:hammodel}
    \widetilde{\mathbf{h}}' = \widetilde{\beta}_0 + \widetilde{\beta}_1 \mathbf{p}'
\end{equation}
Here $\widetilde{\beta}_0$ has size $M^2 \times 1$, while $\widetilde{\beta}_1$ has maximal size $M^2 \times M^2$.  For the smaller molecules in our study (\htwo ~and \heh), where the STO-3G basis set leads to a dimension of $M = 2$, we use (\ref{eqn:hammodel}) with no modifications.

For \lih, a larger molecule, to handle entries of $\mathbf{p}'$ that are identically zero, and also to dramatically reduce the computational effort required for training, we modify the basic model (\ref{eqn:hammodel}).  We can understand these modifications very succinctly by saying that we reduce both the number of columns of $\widetilde{\beta}_1$ and the number of rows of all vectors in (\ref{eqn:hammodel}), namely $\widetilde{\mathbf{h}}'$, $\widetilde{\beta}_0$, and $\mathbf{p}'$.  In more detail, what we do is first form a set $Z$ consisting of indices of, separately, the real and imaginary parts of training data matrices $\{ \mathbf{P}_R'(t_j) \}_{j=0}^N$ and $\{ \mathbf{P}_I'(t_j) \}_{j=0}^N$ that are \emph{not} identically zero.  Let us use the notation  $\widetilde{\mathbf{H}}'$ to denote the $M \times M$ Hermitian matrix that corresponds to the real vector $\widetilde{\mathbf{h}}'$.  For both $\mathbf{P}'$ \emph{and also for our model Hamiltonian} $\widetilde{\mathbf{H}}'$, we restrict attention to upper-triangular matrix indices that are in the set $Z$.  To illustrate this concretely, $\lih$ in the STO-3G basis has dimension $M = 6$, and so, at each instant of time, both $\mathbf{P}'$ and $\widetilde{\mathbf{H}}'$ are determined by $M^2 = 36$ entries, of which $21$ correspond to real parts and $15$ correspond to imaginary parts.  Of these, only $10$ real parts and $6$ imaginary parts are not identically zero.  In this way, we reduce the overall dimension of (\ref{eqn:hammodel}) from $36$ down to $10+6=16$.  We defer the formal details of this procedure to the Appendix.


Note that (\ref{eqn:hammodel}) is by no means the only possible model.  We have explored higher-order polynomial models that, while remaining linear in the parameters $\widetilde{\beta}$, allow $\widetilde{\mathbf{h}}'$ to depend nonlinearly on $\mathbf{p}'$.  We have also explored models in which $\widetilde{\mathbf{h}}'$ is allowed to depend explicitly on time $t$, including through Fourier terms such as $\sin(\omega t)$ and $\cos(\omega t)$.  None of these choices led to any improvement in validation or test error, so we focus on the linear model (\ref{eqn:hammodel}).

Now that we have (\ref{eqn:hammodel}), we turn our attention to (\ref{eq:TDHF_density}).  Then we use a centered-difference approximation to derive from (\ref{eq:TDHF_density}) the statistical model
\begin{equation}
    \label{eqn:statmodel}
    i \frac{ \mathbf{P'}(t_{j+1}) - \mathbf{P'}(t_{j-1}) }{ 2 \Delta t } 
    =  \bigg [   \widetilde{\mathbf{H}}' (\mathbf{P}'(t_j)) , \mathbf{P'}(t_j) \bigg] + \epsilon_j,
\end{equation}
with $\epsilon_j$ denoting error.  With $\| \mathbf{A} \|_F^2 = \sum_{i, j} A_{ij}^2$, the squared Frobenius norm, we form the sum of squared errors loss function
\begin{multline}
    \label{eqn:sseloss}
    \mathcal{L}( \widetilde{\beta} ) = \sum_{j=1}^{N-1} \biggl\| i \frac{ \mathbf{P'}(t_{j+1}) - \mathbf{P'}(t_{j-1}) }{ 2 \Delta t } \\
    -  \bigg [   \widetilde{\mathbf{H}}' (\mathbf{P}'(t_j)) , \mathbf{P'}(t_j) \bigg] \biggr\|_F^2.
\end{multline}

\begin{figure}
  {\centering \includegraphics[clip,trim=100 150 250 0,width=3in]{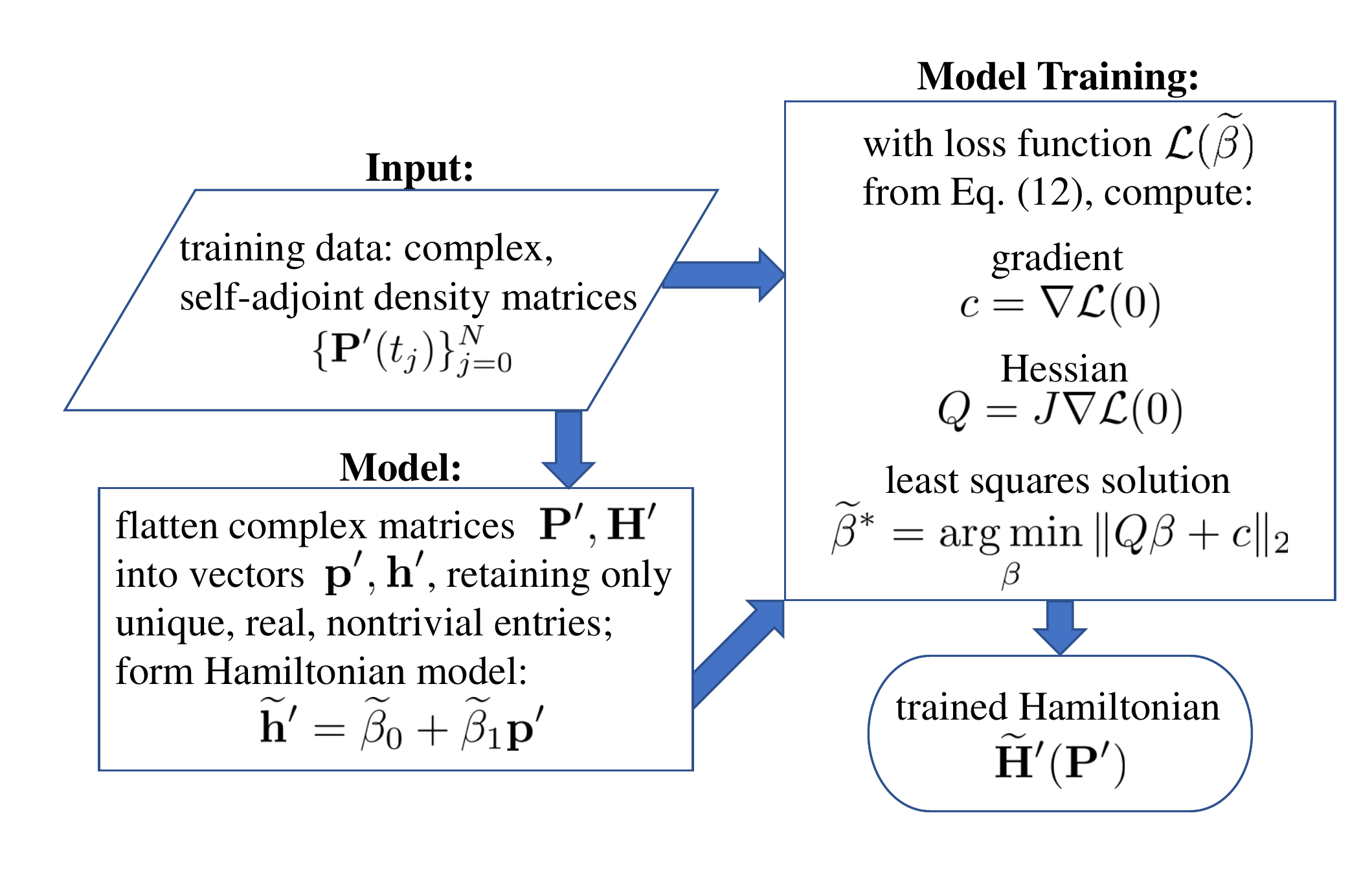}}
  \caption{Overall training procedure for learning the molecular, field-free Hamiltonian.  In this paper, for each molecule, we train using time series with $N=1000$.  We use this field-free Hamiltonian to propagate for $2N=2000$ steps; see Figures \ref{fig:HeH+H2} and \ref{fig:LiH}.  We augment the learned Hamiltonian with an external potential (an electric field), yielding a field-on Hamiltonian that we use to propagate for $2N=2000$ steps; see Figures \ref{fig:HeH+H2WF} and \ref{fig:LiHWF}.}
  \label{fig:flowchart}       
\end{figure}

\subsection{Reduction to Least Squares}
The dependence of $\mathcal{L}$ on $\widetilde{\beta} = ( \widetilde{\beta}_0, \widetilde{\beta}_1 )$ is entirely through $\widetilde{\mathbf{H}}'$.  We estimate $\widetilde{\beta}$ by solving the optimization problem $\widetilde{\beta}^\ast = \argmin_{\beta} \mathcal{L}(\widetilde{\beta})$.  Because (\ref{eqn:hammodel}) is linear in the parameters $\widetilde{\beta}$, we observe that (\ref{eqn:sseloss}) must be quadratic in $\widetilde{\beta}$.  So, there exist constants $Q$ (matrix), $c$ (vector), and $L_0$ (scalar) such that
\begin{equation}
    \label{eqn:quadloss}
\mathcal{L}(\widetilde{\beta}) = \frac{1}{2} \widetilde{\beta}^T Q \widetilde{\beta} + c^T \widetilde{\beta} + \frac{L_0}{2}.
\end{equation}
Here we can identify $c$ as the gradient of $\mathcal{L}$ with respect to $\widetilde{\beta}$ evaluated at $\widetilde{\beta} \equiv 0$, and $Q$ as the Hessian of $\mathcal{L}$ with respect to $\widetilde{\beta}$.  We compute this gradient and Hessian via automatic differentiation of $\mathcal{L}$.
When $Q$ is full rank, we have an exact minimizer $-Q^{-1} c$. 
As $Q$ is typically rank deficient, we replace $Q^{-1}$ with the Moore-Penrose pseudoinverse $Q^{\dagger}$:
\begin{equation}
    \label{eqn:betasol}
    \widetilde{\beta}^\ast = -Q^\dagger c = \argmin_{\beta} \| Q \beta + c \|_2.
\end{equation}
When $(I - Q Q^\dagger) c = 0$, the loss $\mathcal{L}$ achieves its global minimum at $\widetilde{\beta}^\ast$.  For each of our molecules, we find that $\| (I - Q Q^\dagger) c \|$ is small but non-zero.  Still, we find empirically that (\ref{eqn:betasol}) yields a nearly zero-norm gradient of $\mathcal{L}$, as good as what can be achieved via other numerical optimization methods.

We have summarized the overall procedure in flowchart form in Figure \ref{fig:flowchart}.   Equation (\ref{eqn:betasol}) constitutes the end of the training procedure.  In particular, we use a method in NumPy, \texttt{linalg.lstsq}, to compute (\ref{eqn:betasol}), and so we avoid the full computation of $Q^\dagger$.  Note that gradient-based optimization can also be used to minimize the loss (\ref{eqn:sseloss}), as we have verified.  However, such a procedure requires millions of small steps, resulting in a training time (for \lih) that is $500$-$1000$ times larger than the time required by a least-squares solver.

\subsection{Error Metrics}
Inserting (\ref{eqn:betasol}) into (\ref{eqn:quadloss}) and using properties of the pseudoinverse, $(Q^\dagger)^T = (Q^T)^\dagger = Q^\dagger$ together with $Q^\dagger Q Q^\dagger = Q^\dagger$, we obtain the training error
\[
\mathcal{L}(\widetilde{\beta}^\ast) = -\frac{1}{2} \left[ c^T Q^\dagger c + L_0 \right],
\]
the value of the loss function at the optimal set of parameters.  The training error measures a local-in-time error, essentially equivalent to starting at the training data point $\mathbf{P}'(t_j)$, propagating one step forward in time with our learned Hamiltonian (\ref{eqn:hammodel}) and comparing with the very next training data point $\mathbf{P}'(t_{j+1})$.  Aggregating these \emph{one-step errors}---squaring and summing their magnitudes---yields the training error $\mathcal{L}(\widetilde{\beta}^\ast)$.


We contrast the training error with the \emph{propagation error}.  Once we have solved for the optimal parameter values $\widetilde{\beta}^\ast$, the model Hamiltonian (\ref{eqn:hammodel}) is completely determined.  Using this estimated Hamiltonian with the initial condition $\mathbf{P}'(0)$ from our training time series, we solve (\ref{eq:TDHF_density}) forward in time using a Runge-Kutta scheme, generating our statistical estimates of $\widetilde{\mathbf{P}}'(t_j)$ from $j=1$ up to $j=2N=2000$, twice the length of the training data.  For the Runge-Kutta integration, we set absolute and relative tolerances to $10^{-12}$.  We then define the propagation error to be
\begin{equation}
    \label{eqn:properrML}
    \mathcal{E} = \frac{1}{2N} \sum_{j=1}^{2N} \left\| \mathbf{P}(t_j) - \widetilde{\mathbf{P}}(t_j) \right\|_F.
\end{equation}
In contrast to the training error, (\ref{eqn:properrML}) measures the divergence between two trajectories---$\mathbf{P}$ (training) and $\widetilde{\mathbf{P}}$ (propagation of ML Hamiltonian)---over \emph{many} time steps.  Both trajectories have exactly the same initial condition, and hence $j=0$ is excluded from the sum.  For $j > 0$, the two trajectories are computed using different numerical schemes (modified midpoint for the training data and Runge-Kutta for the ML Hamiltonian propagation) \emph{and} different Hamiltonians.  To control for scheme-related error, we compute
\begin{equation}
    \label{eqn:properrEX}
    \mathcal{E}_{\text{Sch}} = \frac{1}{2N} \sum_{j=1}^{2N} \left\| \mathbf{P}(t_j) - \overline{\mathbf{P}}(t_j) \right\|_F,
\end{equation}
where $\overline{\mathbf{P}}(t_j)$ is the result of propagating forward in time using the same Runge-Kutta scheme with the exact Hamiltonian $\mathbf{H}'$.  This exact Hamiltonian is built by (i) extracting the Hamiltonian $\mathbf{H}$ in the AO basis from the electronic structure output and then (ii) transforming $\mathbf{H}$ to $\mathbf{H}'$ using the procedure described in the Appendix.  In (\ref{eqn:properrEX}), the two trajectories being compared have the same Hamiltonian and differ only in the numerical propagation schemes used to generate them.  As a final error metric, we compute
\begin{equation}
    \label{eqn:properr}
    \mathcal{E}_{\text{Ham}} = \frac{1}{2N} \sum_{j=1}^{2N} \left\| \widetilde{\mathbf{P}}(t_j) - \overline{\mathbf{P}}(t_j) \right\|_F.
\end{equation}
The two trajectories compared here are computed using the same Runge-Kutta scheme, but with different Hamiltonians.  By the triangle inequality, we have $\mathcal{E} \leq \mathcal{E}_{\text{Sch}} + \mathcal{E}_{\text{Ham}}$. We may conceptualize this as breaking down the total error into the error due to different schemes ($\mathcal{E}_{\text{Sch}}$) and the error due to different Hamiltonians ($\mathcal{E}_{\text{Ham}}$).

\section{Results}

\setlength{\tabcolsep}{2pt}
\def\arraystretch{1.1}
\begin{table} \centering
\begin{tabular}{r|ccc}
 & \heh & \htwo & \lih \\ \hline 
\rule{0pt}{3ex} $\mathcal{L}(\widetilde{\beta}^\ast)$  & $4.75 \times 10^{-6}$ & $5.77 \times 10^{-6}$ & $2.30 \times 10^{-6}$ \\ 
$\| \nabla \mathcal{L}(\widetilde{\beta}^\ast) \|$  & $4.17 \times 10^{-11}$ & $3.44 \times 10^{-11}$ & $6.47 \times 10^{-11}$ \\ 
$\mathcal{E}$ & $4.37 \times 10^{-3}$ & $4.89 \times 10^{-3}$ & $6.51 \times 10^{-3}$ \\
$\mathcal{E}_{\text{Sch}}$ & $2.57 \times 10^{-3}$ & $2.50 \times 10^{-3}$  & $2.15 \times 10^{-3}$ \\
$\mathcal{E}_{\text{Ham}}$ & $1.81 \times 10^{-3}$ & $2.40 \times 10^{-3}$ & $5.41 \times 10^{-3}$ 
\end{tabular}
\caption{After training, we report the training loss and the norm of its gradient, along with three forms of propagation error. All results are for the field-free problem.
Note that the training error is a sum of squared errors; for each molecule, if we divide by the training data length $N = 10^3$, we obtain mean-squared training errors that are all on the order of $10^{-9}$, indicating approximately $4$ decimal places of accuracy.  The propagation errors show a roughly even breakdown into error due to different schemes versus error due to different Hamiltonians.}
\label{tab:err}
\end{table}

\begin{figure*}
  {\centering \includegraphics[width=3.25in,trim=0 50 0 0,clip]{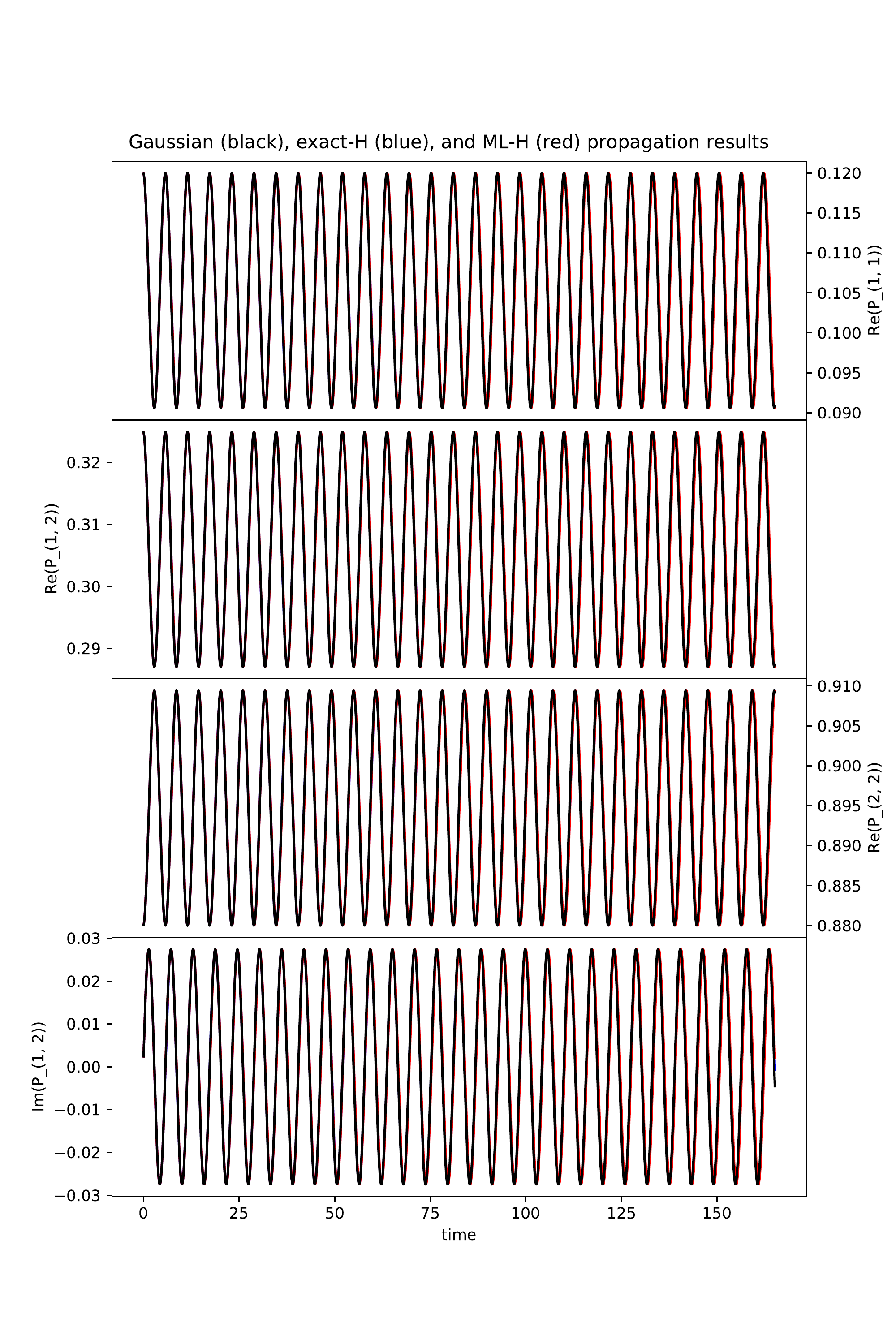}
  \includegraphics[width=3.25in,trim=0 50 0 0,clip]{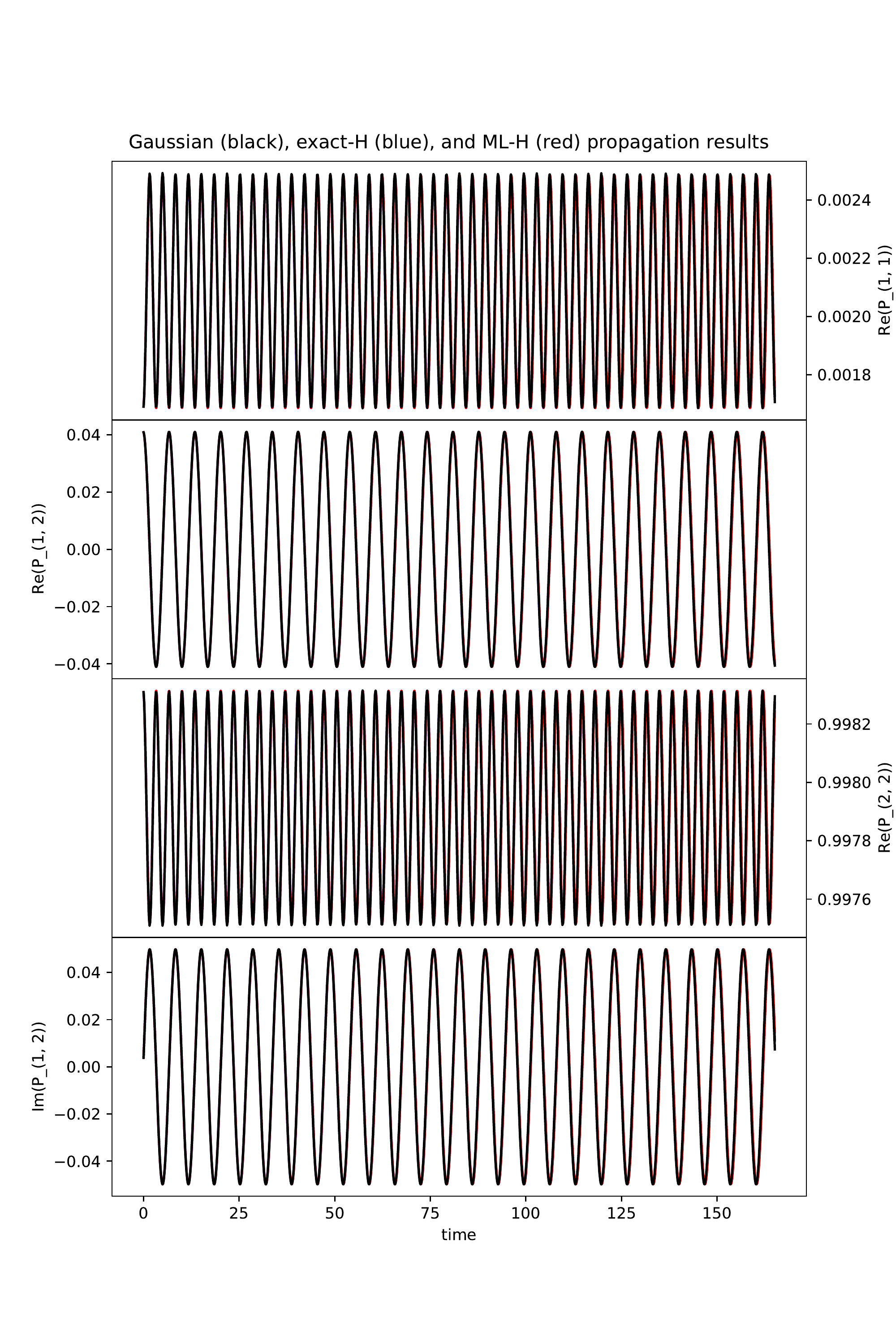} }
  \caption{\heh \, (left) and \htwo \, (right) propagation with no field.   For both molecules, we have plotted all unique real and imaginary parts of the time-dependent density matrices: actual training data (black), exact Hamiltonian propagation (blue), and ML Hamiltonian propagation (red).  Note the close agreement of all three curves, on a time interval that is twice the length used for training.}
  \label{fig:HeH+H2}       
\end{figure*}
\begin{figure}
  {\centering \includegraphics[width=3.25in,trim = 0 80 0 100, clip]{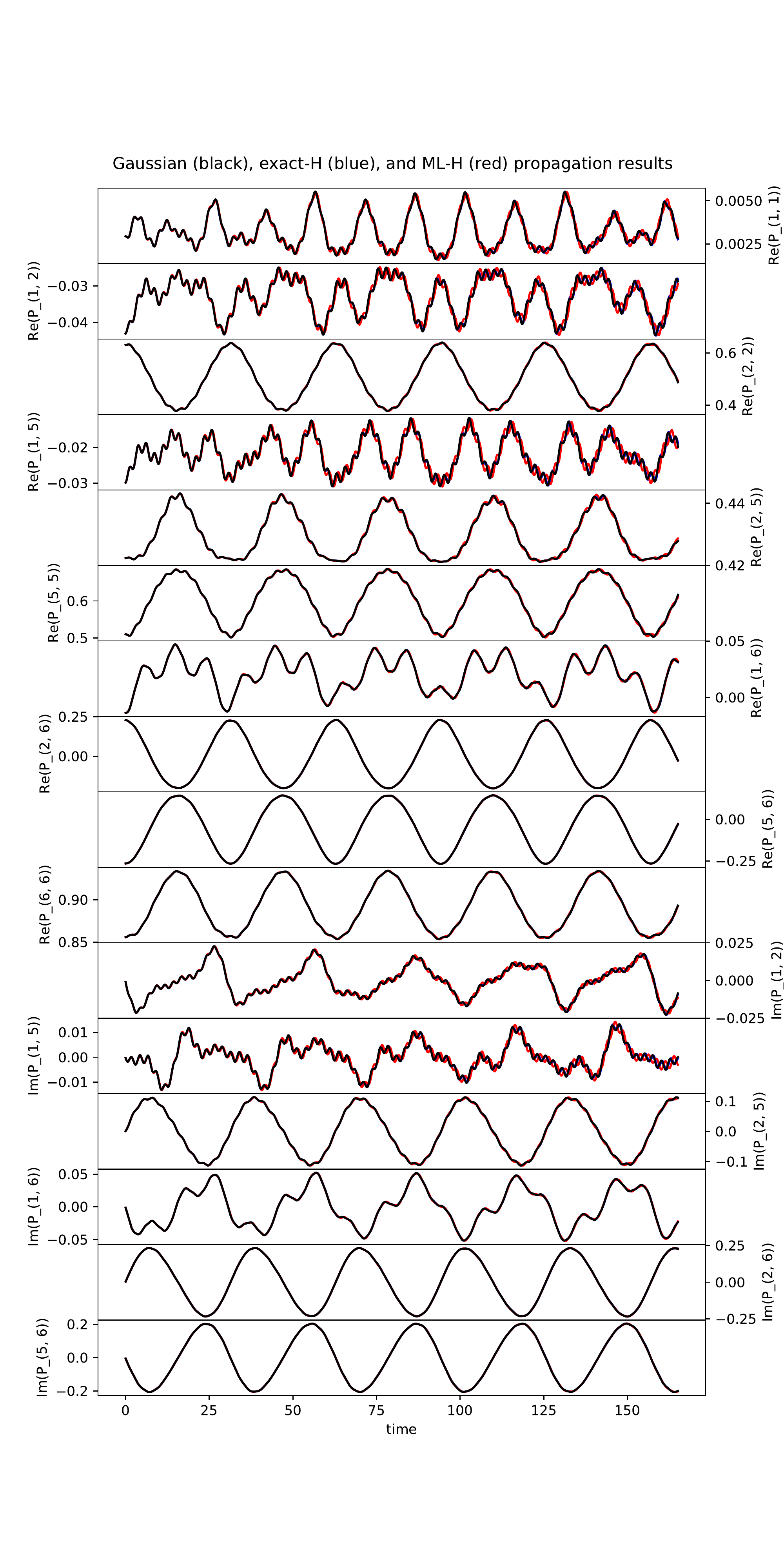} }
  \caption{\lih \ propagation with no field.   We have plotted all unique real and imaginary parts of the time-dependent density matrices: actual training data (black), exact Hamiltonian propagation (blue), and ML Hamiltonian propagation (red).  For density matrix elements with small variance, we discern slight disagreement especially at large times.  For large-variance density matrix elements, the curves are in close agreement.}
  \label{fig:LiH}       
\end{figure}

\begin{figure*}
    {\centering \includegraphics[width=3.25in]{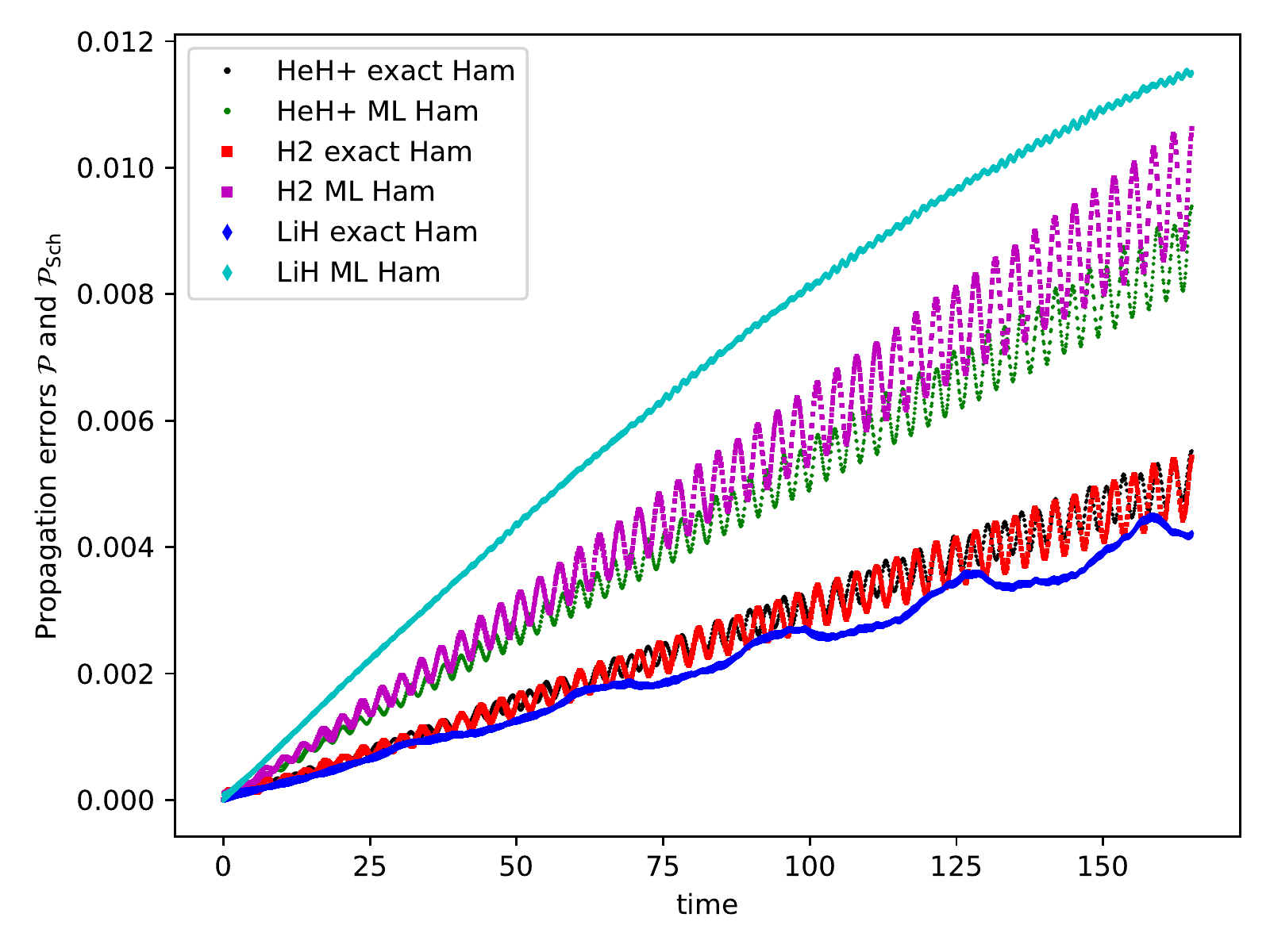}
    \includegraphics[width=3.25in]{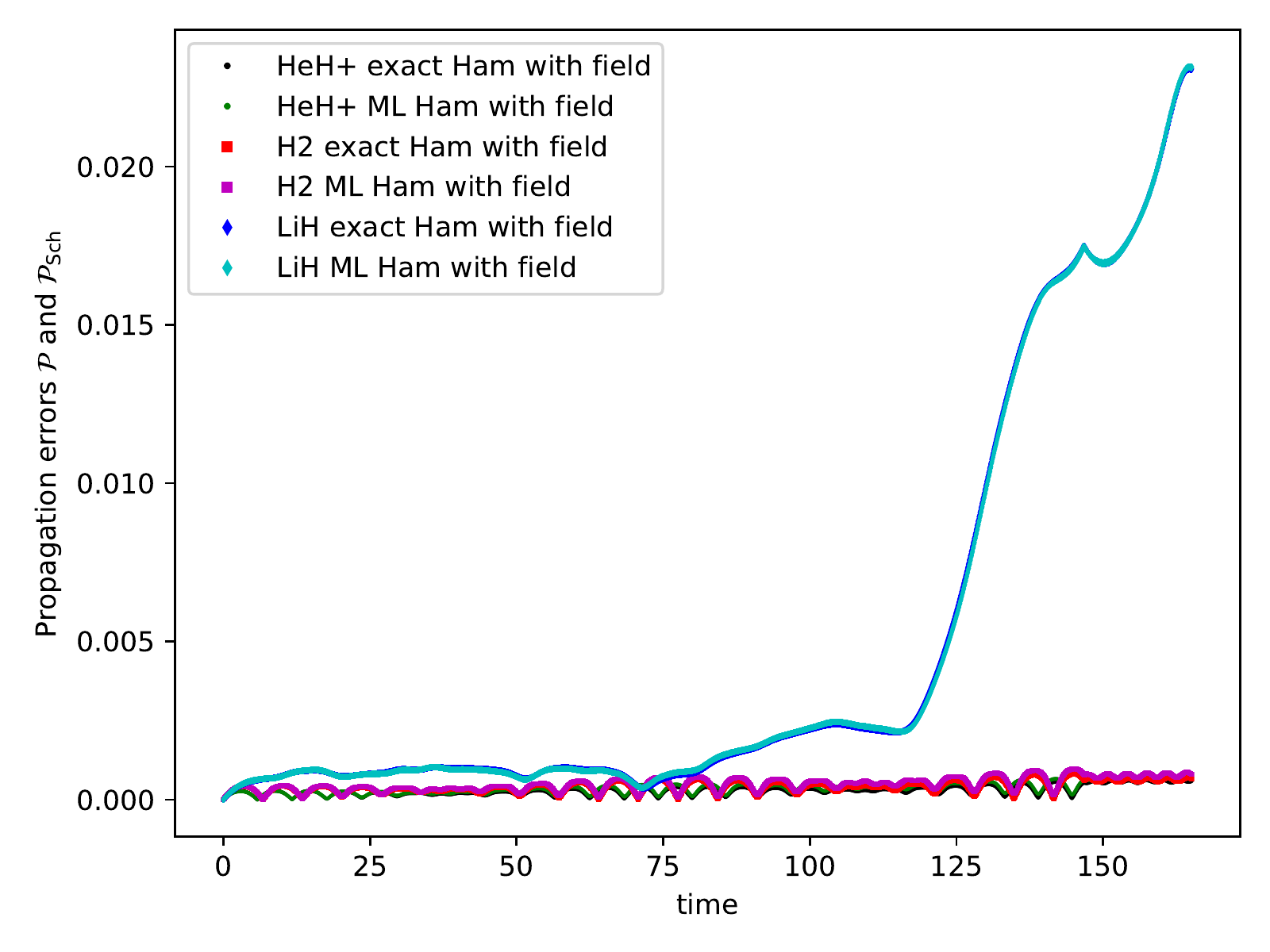} }
    \caption{Time-dependent propagation errors in which we compare the training data against either $\widetilde{\mathbf{P}}$, the result of propagating the ML Hamiltonian, or $\overline{\mathbf{P}}$, the result of propagating the exact Hamiltonian.  All calculations on the left (respectively, right) are for the field-free (respectively, field-on) problem.  For each molecule, the error incurred by propagating with the ML Hamiltonian is within a constant factor of the error incurred by propagating with the exact Hamiltonian.  At the final time, all errors are on the order of $10^{-3}$, except for the field-on calculations with \lih.  The average values of these curves over all time correspond precisely to $\mathcal{E}$ and $\mathcal{E}_{\text{Sch}}$---see (\ref{eqn:properrML}), (\ref{eqn:properrEX}), and Table \ref{tab:err} for further details.}
    \label{fig:hamerr}
\end{figure*}

\subsection{Training and Propagation Tests}
We apply the procedure described in Section \ref{sec:learnham} to training time series of length $N=1000$ for each of the three molecules \heh, \htwo, and \lih.  See Section \ref{sec:1} for details on the generation of training data.  The only additional preprocessing step here was to omit the first two time steps, for each molecule, and to take the subsequent $1000$ time steps as training data.  This was carried out purely to avoid large numerical time-derivatives $d \mathbf{P}'/d t$ associated with the delta-kick perturbation at $t=0$; these time-derivatives form a critical part of our loss function (\ref{eqn:sseloss}).  We emphasize that these training trajectories were generated with no external potential/field, using delta-kick initial conditions described in Section \ref{sect:initcond}.

We report the value of the loss and the norm of its gradient, after training, in the first two rows of Table \ref{tab:err}. For each molecule, the training loss is of the order of $10^{-6}$, which corresponds to an accuracy of roughly 4 decimal places.  In order to visualize this accuracy, see Figures \ref{fig:HeH+H2} and \ref{fig:LiH}.  For each molecule, we have plotted each of the non-zero real and imaginary components (note the $y$-axis labels) that fully determine the Hermitian density matrices $\mathbf{P}'(t_j)$ at each time step $t_j = j \Delta t$.  In fact, in each panel, there are three curves: in black, we have plotted the actual training data produced by the electronic structure code; in blue, we have plotted $\overline{\mathbf{P}}(t_j)$, the result of propagating the exact Hamiltonian; and in red, we have plotted $\widetilde{\mathbf{P}}(t_j)$, the result of propagating the ML Hamiltonian.

For \heh \, and \htwo \, (Fig. \ref{fig:HeH+H2}), the curves agree to a degree where they can hardly be distinguished.  As we described above, due to the fact that in HF theory $\mathbf{H'}$ depends on $\mathbf{P'}$, the TDHF equation (\ref{eq:TDHF_density}) is nonlinear, and hence all of these oscillations are nonlinear oscillations.  For \lih \, (Fig. \ref{fig:LiH}), we can discern some divergence between the result of ML Hamiltonian propagation (red) and the other two curves, but only for those density matrix elements with relatively small variance.  The sum of squares loss function (\ref{eqn:sseloss}) is biased in favor of fitting large-variance components; to avoid this, one could modify (\ref{eqn:sseloss}) to include weights that are inversely proportional to density element variances.  The errors in Figure \ref{fig:LiH} consist primarily of oscillations about the black curve; the magnitudes of these oscillations are small and do not increase dramatically over time.   Still, we should give the the machine-learned Hamiltonian credit for performing well when we use it to  propagate for $2N = 2000$ steps, twice the length of the training data used.  This hints at being able to use the machine-learned Hamiltonian to extrapolate beyond the field-free system used for training.

To understand more deeply the different sources of error, we refer to the final three rows of Table \ref{tab:err} together with the left panel of Figure \ref{fig:hamerr}.  We think of $\mathcal{E}$ as the overall RMS error between the training data $\mathbf{P}'$ and our predicted trajectory $\widetilde{\mathbf{P}}$, broken down into two components $\mathcal{E}_\text{sch}$ and $\mathcal{E}_\text{Ham}$ as explained above.  If our goal is to track the training data, we incur errors of the same order of magnitude when we use either the ML Hamiltonian or the exact Hamiltonian. Consistent with Fig. \ref{fig:LiH}, we find the largest gap between exact and ML Hamiltonian propagation for \lih.

\setlength{\tabcolsep}{4pt}
\def\arraystretch{1.1}
\begin{table} \centering
\begin{tabular}{r|ccc}
 & \heh & \htwo & \lih \\ \hline 
$\mathcal{E}$ & $3.59 \times 10^{-4}$ & $4.97 \times 10^{-4}$ & $4.86 \times 10^{-3}$ \\
$\mathcal{E}_{\text{Sch}}$ & $2.94 \times 10^{-4}$ & $4.10 \times 10^{-4}$  & $4.87 \times 10^{-3}$ \\
$\mathcal{E}_{\text{Ham}}$ & $7.22 \times 10^{-5}$ & $1.01 \times 10^{-4}$ & $1.33 \times 10^{-4}$ 
\end{tabular}
\caption{For the field-on problem, we report three forms of propagation error corresponding to field-on versions of (\ref{eqn:properrML}), (\ref{eqn:properrEX}), and (\ref{eqn:properr}).  Here $\mathcal{E}$ measures the difference between (i) propagation of the ML Hamiltonian plus $\mathbf{V}_{\text{ext}}$ and (ii) the output of an electronic structure code for the field-on problem; $\mathcal{E}_\text{Sch}$ measures the difference between (ii) and (iii) propagation of the exact Hamiltonian plus $\mathbf{V}_{\text{ext}}$.  Finally, $\mathcal{E}_\text{Ham}$ measures the difference between (i) and (iii).  Overall, we find that the errors are lower than in Table \ref{tab:err}, indicating that the ML Hamiltonian succeeds in solving the field-on problem.}
\label{tab:errWF}
\end{table}

\begin{figure*}
  {\centering \includegraphics[width=3.25in,trim=0 20 0 0,clip]{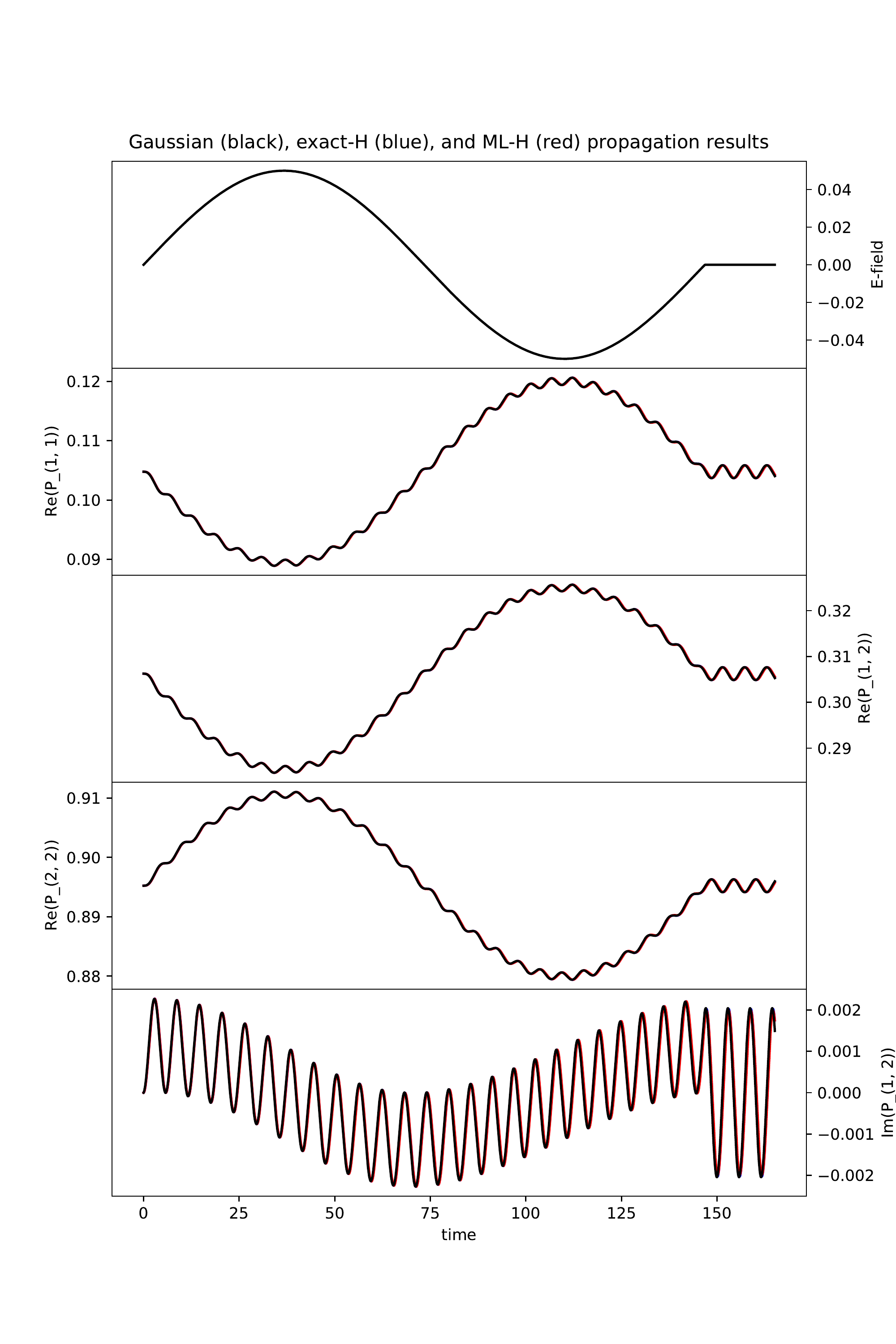}
  \includegraphics[width=3.25in,trim=0 20 0 0,clip]{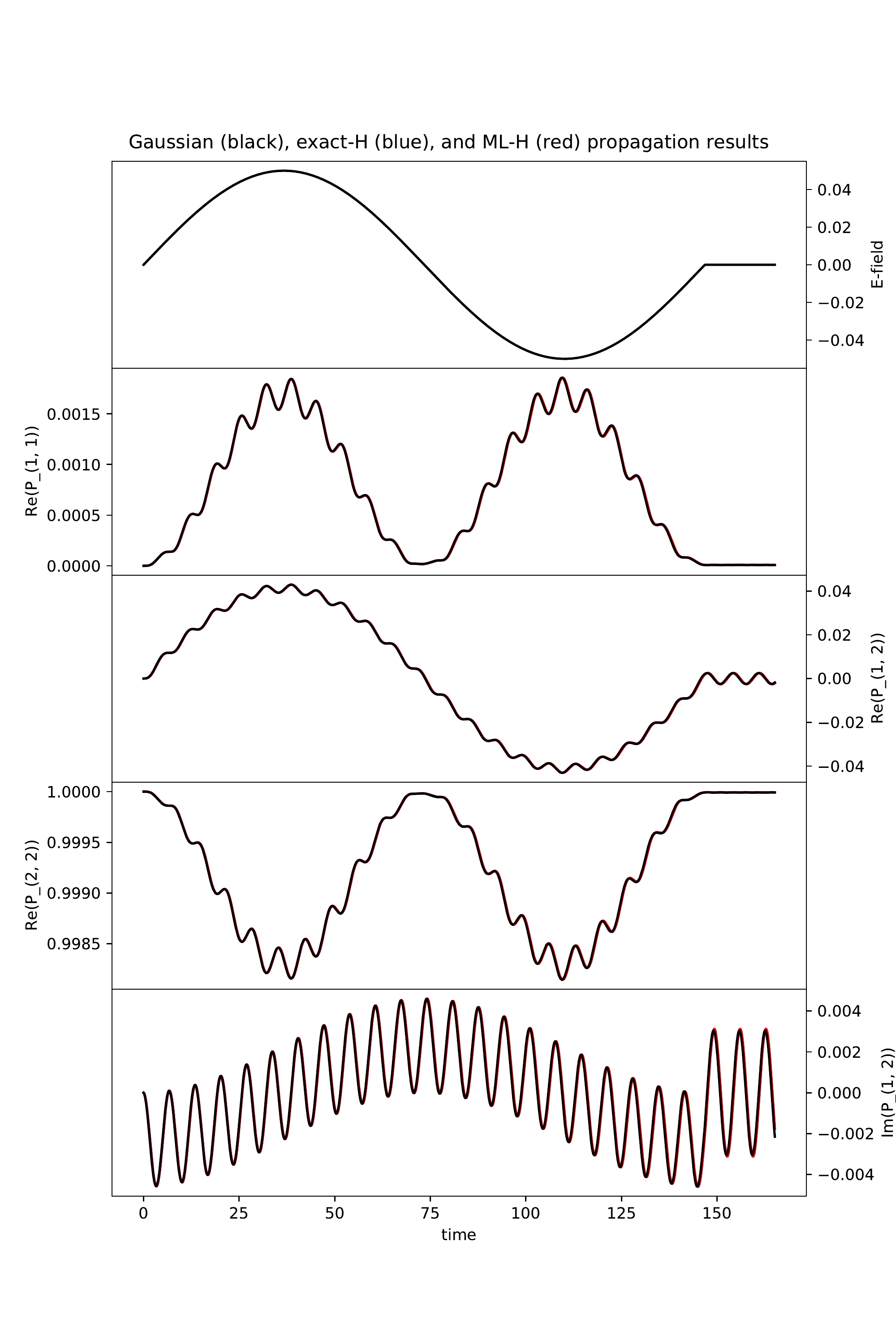} }
  \caption{\heh \, (left) and \htwo \, (right) propagation with field.  The top panel of each plot gives the applied electric field (\ref{eqn:Efield}).  In subsequent panels, for both molecules, we plot all unique real and imaginary parts of the time-dependent density matrices: actual training data (black), exact Hamiltonian propagation (blue), and ML Hamiltonian propagation (red).  By ML Hamiltonian, we mean the Hamiltonian trained on the field-free data plus $\mathbf{V}_{\mathrm{ext}}$ given by (\ref{eqn:Efield}).  Note the close agreement of all three curves, on a time interval that is twice the length used for training.  This is a true test of whether the learned Hamiltonian can extrapolate to problem settings beyond the one used for training.}
  \label{fig:HeH+H2WF}       
\end{figure*}

\begin{figure}
  {\centering \includegraphics[width=3.25in,trim = 0 80 0 100, clip]{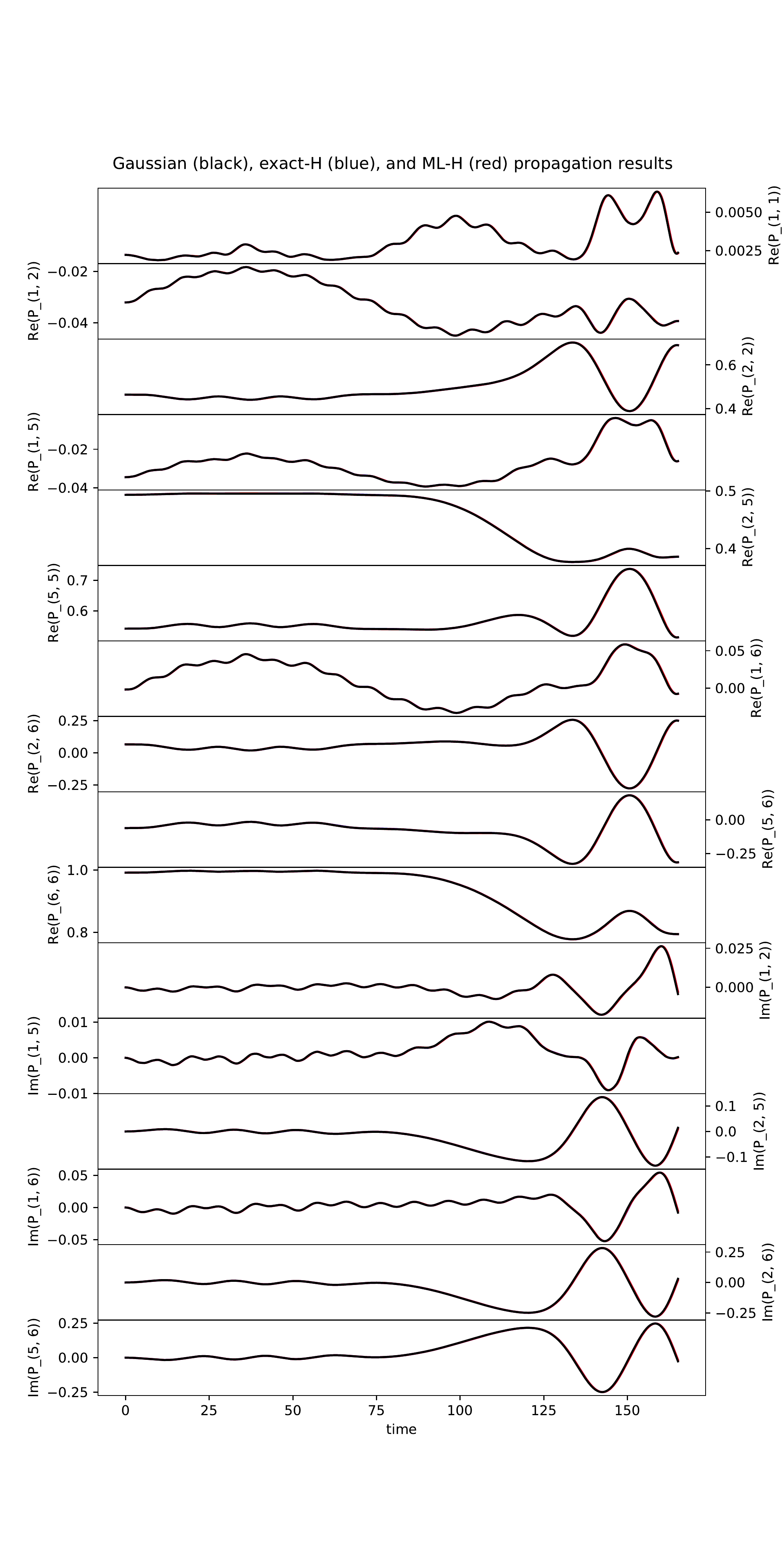} }
  \caption{\lih \ propagation with field.  We plot all unique real and imaginary parts of the time-dependent density matrices: actual training data (black), exact Hamiltonian propagation (blue), and ML Hamiltonian propagation (red).  By ML Hamiltonian, we mean the Hamiltonian trained on the field-free \lih \, data plus $\mathbf{V}_{\mathrm{ext}}$ given by (\ref{eqn:Efield}).  Note the  close agreement of all curves, on a time interval that is twice the length used for training.  This is a true test of whether the learned Hamiltonian can extrapolate to problem settings beyond the one used for training.  We omit a plot of the electric field here---see the top panels of Figure \ref{fig:HeH+H2WF}.}
  \label{fig:LiHWF}       
\end{figure}

\begin{figure}
    \includegraphics[width=3.25in]{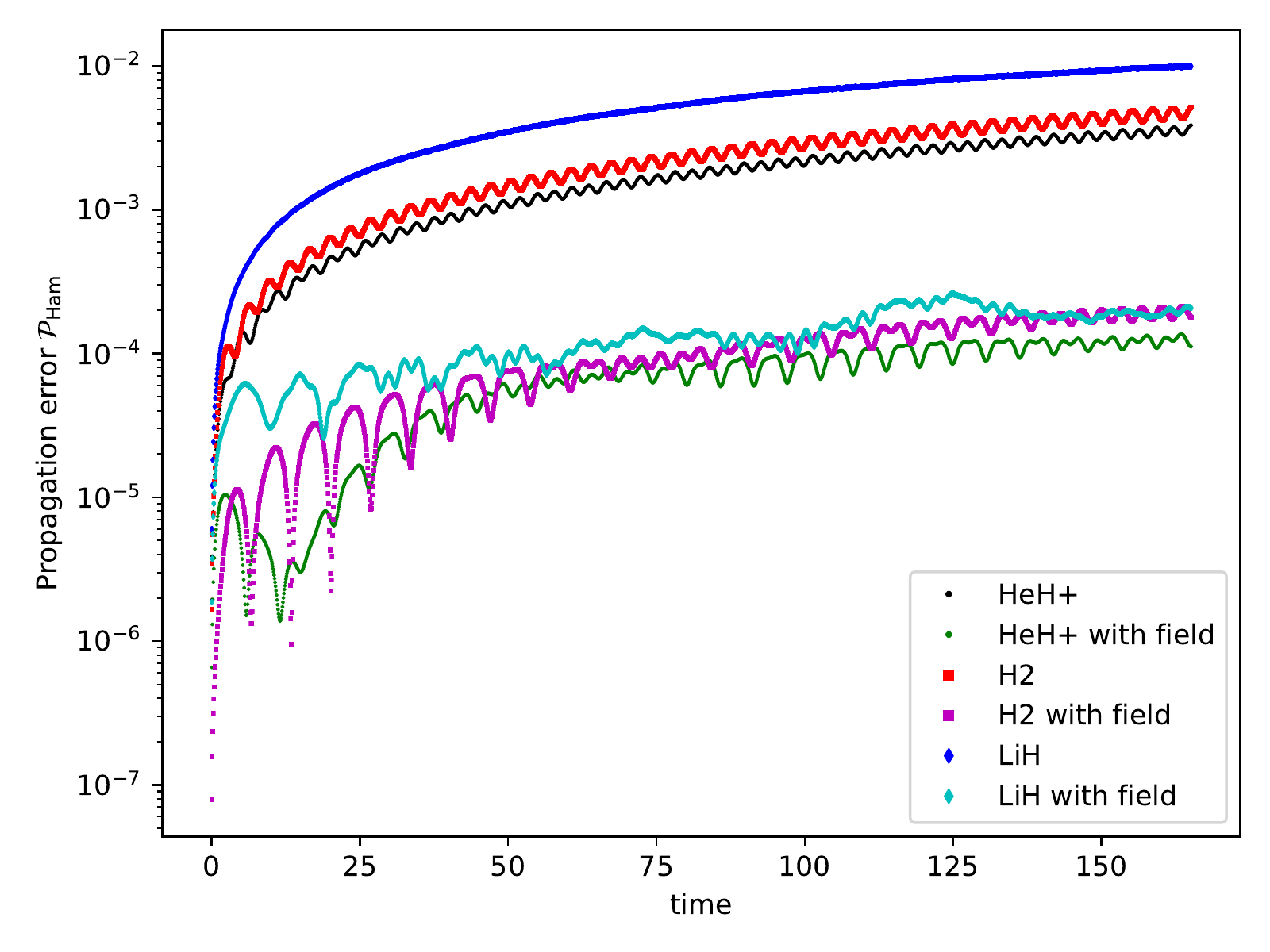}
    \caption{Time-dependent propagation errors in which we compare $\widetilde{\mathbf{P}}$, the result of propagating the ML Hamiltonian, with $\overline{\mathbf{P}}$, the result of propagating the exact Hamiltonian.  All results were computed using the same Runge-Kutta scheme, isolating the error due to the different Hamiltonians.  We include both field-free and field-on calculations.  Note that all results are plotted on a log scale.  The results show that when we propagate both the ML and exact Hamiltonians using the same scheme, the errors between the two resulting trajectories remain small even as we take hundreds of time steps.  The average values of these curves over all time correspond precisely to $\mathcal{E}_{\text{Ham}}$---see (\ref{eqn:properr}) and Table \ref{tab:err} for further details.}
    \label{fig:exmlproperr}
\end{figure}

\subsection{Electric Field Tests}
After learning a field-free Hamiltonian for each of the three molecules, we compared the values of $\mathbf{H}'(t)$ and $\widetilde{\mathbf{H}}'(t)$ along the training trajectories.  We found that the ML Hamiltonian does not equal the exact Hamiltonian.  This led us to question whether the ML Hamiltonian could solve a problem  outside the training regime.  We therefore augmented the ML Hamiltonian with an applied electric field, i.e., the time-dependent external potential $\mathbf{V}_{\mathrm{ext}}$ given in (\ref{eqn:Efield}).  Using the same Runge-Kutta scheme and tolerances described earlier, we propagated for $2N = 2000$ steps.  We compared these results with test data produced by an electronic structure code, and also the results of propagating the exact Hamiltonian, augmented with $\mathbf{V}_{\mathrm{ext}}$, via our Runge-Kutta method.


For a first view of the field-on results, see Table \ref{tab:errWF} and Figures \ref{fig:HeH+H2WF} and \ref{fig:LiHWF}.  In particular, the top panels of Figure \ref{fig:HeH+H2WF} show the applied electric field; note that it is switched off abruptly after one period.  We can immediately discern that the applied field substantially alters the electron density from the field-off case.  Still, in each panel, we see excellent agreement between all three curves in each plot: the ground truth solution produced by an electronic structure code (black), the result of propagating the exact Hamiltonian plus $\mathbf{V}_{\text{ext}}$ (blue), and the result of propagating the ML Hamiltonian plus $\mathbf{V}_{\text{ext}}$ (red).  Table \ref{tab:errWF}, in which we find errors that are roughly an order of magnitude lower than those in Table \ref{tab:err}, confirms that all computed densities are in close quantitative agreement.  To repeat, all field-on results are for a time interval that is twice the length used for training, and training was conducted using field-off data only. Overall, we take these results to indicate that the ML Hamiltonian can indeed extrapolate to problem settings beyond the one used for training.

For a deeper understanding of the field-on results, we focus on the right panel of Figure \ref{fig:hamerr} and Figure \ref{fig:exmlproperr}.   In the right panel of Figure \ref{fig:hamerr}, we compare (i) the result of propagating the ML Hamiltonian plus $\mathbf{V}_{\mathrm{ext}}$ against (ii) the ground truth solution, the output of the electronic structure code for the field-on problem.  We also compare (ii) with (iii) the result of propagating the exact Hamiltonian plus $\mathbf{V}_{\mathrm{ext}}$. The plots indicate that, for all three molecules and especially for \lih, the error between (i) and (ii) is almost identical to that between (ii) and (iii).  This indicates that the bulk of the error is due to our use of a Runge-Kutta scheme instead of the MMUT scheme used in the electronic structure code.  To confirm this, we consult Figure \ref{fig:exmlproperr}, in which we compare (i) and (iii) directly.  All solutions here are computed using the same Runge-Kutta scheme.  For each molecule, we see that the errors for the field-on problems are consistently smaller than those for the field-off problems.  We conclude from these results that the ML Hamiltonian can be used to compute the electronic response to an applied electric field.

A short derivation will show that it is not automatic to expect the augmented ML Hamiltonian to propagate correctly.  Let us work in continuous time, to eliminate error due to discrete-time propagation; in this idealized setting, we start with the statement that both of our field-free Hamiltonians, $\mathbf{H}'(t)$ (exact) and $\widetilde{\mathbf{H}}'(t)$ (ML), satisfy the TDHF equation:
\begin{align*}
i \frac{d \mathbf{P}'(t)}{d t} &= [ \mathbf{H}'(t), \mathbf{P}'(t) ] \\
i \frac{d \mathbf{P}'(t)}{d t} &= [ \widetilde{\mathbf{H}}'(t), \mathbf{P}'(t) ].
\end{align*}
Subtracting these equations, and defining the error $\boldsymbol \epsilon(t) = \mathbf{H}'(t) - \widetilde{\mathbf{H}}'(t)$, we obtain
\begin{equation}
\label{eqn:epscomm}
[ \boldsymbol \epsilon(t), \mathbf{P}'(t) ] = 0.
\end{equation}
Now we augment both Hamiltonians with an external field $\mathbf{V}_{\text{ext}}(t)$.  Let $\mathbf{P}''(t)$ denote the true density for the problem with the external field.  It must satisfy
\[
i \frac{d \mathbf{P}''(t)}{d t} = [\mathbf{H}'(t) + \mathbf{V}_{\text{ext}}(t), \mathbf{P}''(t) ].
\]
Via $\mathbf{H}'(t) = \boldsymbol \epsilon(t) + \widetilde{\mathbf{H}}'(t)$, we obtain
\[
i \frac{d \mathbf{P}''(t)}{d t} = [\widetilde{\mathbf{H}}'(t) + \mathbf{V}_{\text{ext}}(t), \mathbf{P}''(t) ] + \underbrace{[\boldsymbol \epsilon(t), \mathbf{P}''(t)]}_{\ast}.
\]
As (\ref{eqn:epscomm}) does not in general imply that the starred term vanishes, we cannot conclude that the true density $\mathbf{P}''(t)$ satisfies the TDHF equation with the augmented ML Hamiltonian $\widetilde{\mathbf{H}}'(t) + \mathbf{V}_{\text{ext}}(t)$.  Based on the above derivation, if we solve the TDHF equation using the augmented ML Hamiltonian, we  expect to obtain a time-dependent density that differs from $\mathbf{P}''(t)$.  As we are able to use the ML Hamiltonian successfully on the problem with an applied electric field, we hypothesize that the error $\boldsymbol \epsilon(t)$ is structured in such a way that enables us to extrapolate to new problems.  We plan to test this hypothesis in future work.

\subsection{Reproducibility}
All code required to reproduce all training and test results (including plots) is available on GitHub at \url{https://github.com/hbhat4000/electrondynamics}. Training data is available from the authors upon request.

\section{Discussion}
Our current work demonstrates that, from a single time series consisting of time-dependent density matrices, we can effectively learn an integrated Hamiltonian matrix.  This ML Hamiltonian can be used for propagation in both the field-off and field-on settings. Importantly, training with a single field-free trajectory, our ML Hamiltonian has the potential to predict the electronic response to a large variety of field pulse perturbations, opening the door to laser-field controlled chemistry.  
The present work leads to two main areas of future work.  The first area concerns technical improvements to the procedure itself, including (i) to replace (\ref{eqn:sseloss}) with a weighted loss function, to account for density elements that oscillate on different vertical scales, (ii) to propagate our ML Hamiltonian using the MMUT scheme, thus eliminating the kind of error quantified by $\mathcal{E}_{\text{Sch}}$, and (iii) to further explore reducing the number of degrees of freedom in the ML Hamiltonian.  The second area concerns improving our overall understanding of the procedure, and applying it to systems of greater chemical and physical interest.  In this area, further work is needed to understand the difference between the exact and ML Hamiltonians, whether this difference can be decreased by training on multiple trajectories, and how far outside the training regime we can push the ML Hamiltonian.  We can also seek to learn the $\hat{H}$ operator rather than the $\mathbf{H}$ matrix representation, which is of interest for determining the unknown exchange-correlation potential within time-dependent density functional theory.  In this way, we can push this procedure beyond known physics (as explored here) to systems where the underlying potential energy terms are not known with sufficient accuracy or precision.

\begin{acknowledgements}
This work was supported by the U.S. Department of Energy, Office of Science, Office of Basic Energy Sciences under Award Number DE-SC0020203.  We acknowledge computational time on the MERCED cluster (funded by NSF ACI-1429783), and on the Nautilus cluster, which is supported by the Pacific Research Platform (NSF ACI-1541349), CHASE-CI (NSF CNS-1730158), and Towards a National Research Platform (NSF OAC-1826967). Additional funding for Nautilus has been supplied by the University of California Office of the President.
\end{acknowledgements}

%
%

\section*{Appendix}
\paragraph{Canonical Orthogonalization.} 
Let $\mathbf{S}$ be the overlap matrix with $ S_{\mu \nu} = \langle \chi_{\mu} | \chi_{\nu} \rangle$.  Because it is real and symmetric, we have $\mathbf{S} = \mathbf{U} \mathbf{s} \mathbf{U}^T$
where $\mathbf{s}$ is diagonal and real, and $\mathbf{U}$ is a real orthogonal matrix.  Now we form $\mathbf{X} = \mathbf{U} \mathbf{s}^{-1/2}$.  Then, we go from $\mathbf{P}$ to $\mathbf{P}'$ via
\[
\mathbf{P}' = \mathbf{s}^{1/2} \mathbf{U}^T \mathbf{P} \mathbf{U} \mathbf{s}^{1/2}.
\]
If $\mathbf{H}$ is the Hamiltonian in the AO basis, the Hamiltonian in the orthogonalized basis is
\[
\mathbf{H}' = \mathbf{s}^{-1/2} \mathbf{U}^T \mathbf{H} \mathbf{U} \mathbf{s}^{-1/2}.
\]
\paragraph{Dimensionality Reduction.} For LiH, certain elements of the density matrix $\mathbf{P}'(t)$ are identically zero for all $t$. We thus define a reduced state vector $\mathbf{p}^\sharp$ that consists of the non-zero upper-triangular degrees of freedom, i.e., the $\widetilde{M} \leq M^2$ elements that are necessary to reconstruct all of $\mathbf{P}'$. Out of these $\widetilde{M}$ elements, we take the first $\widetilde{M}_R$ to correspond to elements of $\mathbf{P}'_R$ and the remaining $\widetilde{M}_I = \widetilde{M} - \widetilde{M}_R$ to correspond to elements of $\mathbf{P}'_I$.  Define $Z$ by
\begin{subequations}
\label{eqn:Zdef}
\begin{gather}
Z = Z^R \cup Z^I \\
Z^R = \{ (i,j) \text{ s.t. } i \leq j \text{ and not } P^R_{ij}(t) \equiv 0 \} \\
Z^I = \{ (i,j) \text{ s.t. } i < j \text{ and not } P^I_{ij}(t) \equiv 0 \}.
\end{gather}
\end{subequations}
We form a mapping $\sigma : \{ 1, 2, \ldots, \widetilde{M} \} \to Z$, whose purpose is to map one-dimensional indices of the reduced vector $\mathbf{p}^\sharp$ to ordered pair indices of the full matrix $\mathbf{P}'$.  We define $\sigma$ implicitly through (\ref{eqn:Zdef}) and the following:
\begin{equation}
\label{eqn:psharp}
p^\sharp_k = \begin{cases} {\mathbf{P}'_R}_{\sigma(k)} & k \leq \widetilde{M}_R \\
 {\mathbf{P}'_I}_{\sigma(k)} & k > \widetilde{M}_R. \end{cases}
\end{equation}
Looping over the entries $k = 1, 2, \ldots, \widetilde{M}$, this equation let us go back and forth from the full complex matrix $\mathbf{P}'$ to the reduced real state vector $\mathbf{p}^\sharp$.

Importantly, we now follow precisely the same procedure, with the same mapping $\sigma$ and set $Z$, to form a reduced Hamiltonian vector $\mathbf{h}^\sharp$.  We then formulate a reduced-dimensionality version of (\ref{eqn:hammodel}):
\begin{equation}
    \label{eqn:hammodelsharp}
        \widetilde{\mathbf{h}}^\sharp = \widetilde{\beta}_0 + \widetilde{\beta}_1 \mathbf{p}^\sharp.
\end{equation}
The matrix $\widetilde{\beta}_1$ is now of dimension $\widetilde{M} \times \widetilde{M}$; all other objects in this equation are vectors of dimension $\widetilde{M} \times 1$.  The training procedure then holds without further modifications.


%
%

\bibliographystyle{spmpsci}
\bibliography{Journal_Short_Name,references}	

\end{document}